\documentstyle[prd,aps,preprint,epsfig]{revtex}

\tighten

\begin{document}
\draft
 
\pagestyle{empty}

\preprint{
\noindent
\hfill
\begin{minipage}[t]{3in}
\begin{flushright}
LBNL--42165 \\
August 1998
\end{flushright}
\end{minipage}
}

\title{The final-state interaction phases of the two-body $B$ decay:\\
The bounds from the updated data}

\author{
Mahiko Suzuki
}
\address{
Department of Physics and Lawrence Berkeley National Laboratory\\
University of California, Berkeley, California 94720
}


\maketitle

\begin{abstract}

The updated experimental data are used to analyze the final-state 
interaction phases of the two-body decay amplitudes of the $B$ mesons,
$B\rightarrow \overline{D}\pi$, $\overline{D}\rho$, $\overline{D}^*\pi$, 
and $\overline{D}^*\rho$. Combining the upper bounds on the branching 
fractions of the color-suppressed neutral modes with those of the 
charged modes, we have set constraints on the relative phases between 
the amplitudes $A(B^0\rightarrow X^-Y^+)$ and $A(B^+\rightarrow X^0Y^+)$ 
where $X=\overline{D}$ or $\overline{D}^*$ and $Y=\pi$ or $\rho$. 
The numbers that we have obtained point to small final-state 
interactions. When these relative phases are expressed in 
those of the isospin amplitudes, the bounds become less tight
since the experimental errors accumulate. In the decay where
many multibody channels are open, however, there is little 
advantage in breaking up the observed amplitudes into the 
isospin eigenchannels for analysis of the final-state interactions.
\end{abstract}

\pagestyle{plain}
\narrowtext

\setcounter{footnote}{0}
\section{Introduction}

It is important to know in the nonleptonic decay of the B mesons how much 
phase is generated for the decay amplitudes by the final-state interaction.
Many calculations were made on the short-distance effects assuming that the 
long-distance effects be small or simply ignoring them\cite{small}.
Over the years various arguments have been presented in support of small to 
vanishing long-distance phases for the two-body decay\cite{BJ,Regge,Suzuki}. 
Since there is no method to compute the long-distance effects accurately, 
some warned about the possibility of large phases\cite{large}. Experimentally,
persistence of the color suppression is one strong qualitative evidence 
for the small phases. To learn about the final-state interaction phases of 
the $B$ decay, we have analyzed here the recently updated data\cite{PDG} 
on the the two-body decay modes. Specifically we have chosen the decay modes 
$B\rightarrow \overline{D}\pi, \overline{D}^*\pi, \overline{D}\rho$, and 
$\overline{D}^*\rho$, which proceed through the nonpenguin interactions. 
With the current experimental uncertainties, the data are consistent with 
vanishing dynamical phases in all cases. Thanks to the substantial 
improvement in the accuracy of measurement, however, our analysis sets the 
meaningful upper bounds on the relative phases of the decay amplitudes. 
The most stringent bound has been set at the level of $10^{\circ}$. 

Before starting, we would like to point out significance and insignificance of 
the isospin amplitudes in the nonleptonic decays. In the decays where only a
small number of decay channels are open, analyzing the isospin amplitudes has 
a clear advantage.  In the extreme case where only the state $AB$ and its
isospin-related states are allowed, we should study their isospin eigenstates 
since the decay amplitudes of definite isospin carry the strong interaction 
eigenphases of elastic $AB$ scattering\cite{Watson}. When another final state 
$CD$ exists and couples to $AB$, it still makes sense to analyze the $2\times 
2$ S-matrix of $AB$ and $CD$ with definite isospin. However, the advantage 
disappears when more than a few channels is open and a channel coupling 
occurs in the final state. In this case the strong interaction S-matrix is an 
$N\times N$ matrix ($N\gg 1$). In terms of the eigenphase shifts $\delta_a$ 
defined by $\langle b|S|a\rangle = \delta_{ba}e^{2i\delta_a} (a,b=1,2,3,
\cdots N)$, the decay amplitude into a hadron channel $h (e.g., D^-\pi^+$ or 
an isospin eigenstate of $\overline{D}\pi)$ can be expressed as 
\begin{equation}
    A(B\rightarrow h) = 
            \sum_{a=1,2,\cdots N} A(B\rightarrow a) e^{i\delta_a}O_{ha},
\end{equation}
where $O_{ha}$ is the diagonalization matrix between the hadron basis and 
the eigenchannels:
\begin{equation}
    |h\rangle = \sum_{a}O_{ha}|a\rangle.
\end{equation}
If the decay occurs through the interactions carrying a common CP-phase ({\it 
e.g.,} $\sim(\overline{d}_L\gamma^{\mu}u_L)(\overline{c}_L\gamma_{\mu}b_L)$ 
and the interactions arising from the QCD corrections to it), the CP-phase 
factors out: $A(B\rightarrow a)^* = A(B\rightarrow a)e^{-2i\delta_{CP}}$. 
Unfortunately we have no practical way to solve the multichannel problem for 
$\delta_a$ and $O_{ha}$. If, for instance, the $B^0\rightarrow\overline{D}
\pi$ amplitude of $I=1/2$ involves $N (\gg 1)$ eigenchannels, the $B^0
\rightarrow D^-\pi^+$ amplitude would also contain roughly as many 
eigenchannel amplitudes, only by a factor of two or so more. In neither 
case is the decay phase simply related to the strong-interaction 
eignephases. That is to say, breaking up the two-body states into the 
isospin eigenchannels accomplishes very little in relating the decay phases 
the strong-interaction S-matrix when scattering is highly inelastic. 
Therefore there is no intrinsic merit in studying the phases of the 
$I$=1/2 and 3/2 amplitudes of $B^0\rightarrow\overline{D}\pi$ instead 
of the phases of the $B^0\rightarrow D^-\pi^+$ and $B^0\rightarrow
\overline{D}^0\pi^0$ amplitudes. It would not be surprising if we have already 
encountered this situation in the $D$ decay. The $K^-\pi^+$ channel of the
$D^0$ decay couples to $\overline{K}^0\pi^0, \overline{K}^0\eta$, and several
$\overline{K}\pi\pi\pi$ channels, resonant and nonresonant with different
internal quantum numbers. In the past, analysis was made for the isospin 
amplitudes of $D\rightarrow\overline{K}\pi$ and $K\overline{K}$\cite{Exp}. 
In the presence of many other decay channels open, we may equally well present 
the decay phases for the directly observed amplitudes instead of the isospin 
amplitudes, particularly in the $B$ decay. 

\section{Parametrization of amplitudes and experimental data}

Since our purpose is to learn about how large the final-state interaction 
phases are in the $B$ decay, we wish to separate out a CP-phase from the 
decay amplitudes.  For this reason we consider the decay modes in which the 
nonpenguin interactions dominate. Best measured are the two-body decay 
modes which are caused by the quark process $b \rightarrow c\overline{u}d$. 
We analyze four sets of the two-body decay modes:
\begin{equation}
      \left\{\begin{array}{l}
         B\rightarrow \overline{D}\pi \\
         B\rightarrow \overline{D}\rho \\
         B\rightarrow \overline{D}^*\pi \\
         B\rightarrow \overline{D}^*\rho.
              \end{array}  \right .
\end{equation}
Each set consists of three decay modes, for instance,
$B^+\rightarrow\overline{D}^0\pi^+$, $B^0\rightarrow D^-\pi^+$, and
$B^0\rightarrow\overline{D}^0\pi^0$ for $B\rightarrow\overline{D}\pi$. 
All four sets of decays have the same isospin structure. Since the weak 
hamiltonian transforms like $\Delta I = 1$, there are two independent 
decay amplitudes in each set.  Choosing $B\rightarrow\overline{D}\pi$ as
an example, we can parametrize the observed amplitudes in terms of the isospin
amplitudes as
\begin{eqnarray}
  A_{0+}&\equiv& A(B^+\rightarrow\overline{D}^0\pi^+) =  A_{3/2}\\ \nonumber 
  A_{-+}&\equiv& A(B^0\rightarrow D^-\pi^+) =
                                   \frac{1}{3}(A_{3/2}+2A_{1/2})\\ \nonumber
  A_{00}&\equiv& A(B^0\rightarrow\overline{D}^0\pi^0) = 
                                   \frac{\sqrt{2}}{3}(A_{3/2}-A_{1/2}).
\end{eqnarray}
Then the three amplitudes obey the sum rule,
\begin{equation}
        A_{0+} - A_{-+} = \sqrt{2}A_{00}. \label{sumrule}
\end{equation}
We denote two relative phases as
\begin{eqnarray}
       \delta_{-+} &=& \arg(A_{-+}/A_{0+}),\\ \nonumber
       \delta_{00} &=& \arg(A_{00}/A_{0+}).
\end{eqnarray}
With the constraint of the sum rule Eq.(\ref{sumrule}), the two phases
are dependent. We can use alternatively the phase difference of the isospin 
amplitudes,
\begin{equation}
       \delta_I = \arg(A_{1/2}/A_{3/2})
\end{equation} 
for parametrization.

From the 1998 edition of the Review of Particle Physics\cite{PDG}, 
we have extracted the magnitudes of amplitude after making the phase space 
corrections of $p^{2l+1}$ on the assumption that the $s$-wave dominates in 
$\overline{D}\pi$ and $\overline{D}^*\rho$, and the $p$-wave in 
$\overline{D}^*\pi$ and $\overline{D}\rho$. The results are tabulated in 
Table I where $|A_{-+}|$ is normalized to unity up to experimental
uncertainties. Only upper bounds have been measured for $|A_{00}|$.
We treat the experimental errors for the three amplitudes as uncorrelated. 
Actually a small portion of the errors ($1\pm 0.0128$) in $|A_{-+}|$ 
and $|A_{00}|$ comes from a common source, which is the lifetime of $B^0$. 
However, this hardly affects our final numbers.

\section{Results of analysis}

  The sum rule Eq.(\ref{sumrule}) can be expressed as a triangular 
relation in the complex plane for each set of the decay modes.  
A typical pattern of the triangular relation is depicted in Figure 1, 
where the phase of $A_{0+}$ is chosen to be zero for reference. 
$A_{00}$ is confined inside the circle.  The sign ambiguity or the 
phase ambiguity by $\pi$ of $A_{-+}/A_{0+}$ has been fixed such that 
the three amplitudes be consistent with the sum rule. The sum rule 
has the ambiguity of the reflection with respect to the real axis.
We have fixed this reflection ambiguity or the complex conjugation 
ambiguity by choosing $\delta_{-+}$ 
between $0^{\circ}$ and $180^{\circ}$. Then $\delta_{00}$ is negative 
by the sum rule.  The bands shown by broken curves at the ends of 
the arrows indicate the experimental errors. Note that for $A_{-+}$ 
the arrow is attached to the direction of $-A_{-+}$. In all cases 
the triangular relation can be satisfied with zero phases if we take 
account of the experimental uncertainties. Here we pose the following 
question: Up to how large phases can be accommodated by the current 
data if we take the quoted experimental errors seriously?

   We have tabulated the answer to the question in Table II. Listed 
are the bounds on the relative phases $\delta_{-+}$ and $\delta_{00}$. 
In obtaining those bounds, the quoted experimental uncertainties have 
been taken into account as uncorrelated errors. For comparison, 
we have also listed the corresponding values for the decay 
$D\rightarrow\overline{K}\pi$ which has the identical isospin 
properties as $B\rightarrow\overline{D}\pi$. Its decay interactions
are also the same in structure up to the replacement of $b\rightarrow c$ 
and $c\rightarrow s$.  The most important, albeit anticipated, conclusion 
is that the phase $\delta_{-+}$ between $A_{-+}$ and $A_{0+}$ must be 
small in all cases except possibly for $B^0\rightarrow\overline{D}\pi$. 
As the measurement on the branching fractions, particularly of the 
color-suppressed modes, will improve in the future, either the upper bounds 
listed in Table II will be tightened or actual values may emerge 
for $\delta_{-+}$. We are not far from seeing the actual values. 
In contrast to $\delta_{-+}$, the phase $\delta_{00}$ between $A_{00}$ 
and $A_{0+}$ is only loosely constrained. The reason is fairly obvious 
in Figure 1: Though the triangle is very flat, {\it i.e.,} the final-state 
interaction is small, the smallness of $|A_{00}|$ leaves room for the 
phase of $A_{00}$ to be large. Even if $\delta_{00}$ turns out to be 
large in the future, it should be interpreted as an accident due to 
the smallness of $|A_{00}|$, not as a consequence of large final-state 
interactions.   
 
      We can express the phases in terms of the isospin amplitude 
phase $\delta_I = \delta_{1/2}-\delta_{3/2}$. In the last column 
of Table II we have listed $\delta_I$.\footnote{In the present phase 
convention, the bottom entry for $D\rightarrow\overline{K}\pi$ is
$(97_{-13}^{+12})^{\circ}$ in \cite{Exp} based on the CLEO data
as of 1996 with more generous errors attached.}  Since it is
$\delta_{-+}$ and $\delta_{00}$ that experiment measures directly,
the experimental uncertainties accumulate and get enhanced when 
expressed in $\delta_I$.

It is a clear conclusion of our analysis that the final-state interaction 
is indeed small and the phase must be fairly small at least for 
$B\rightarrow\overline{D}\pi, \overline{D}\rho$, and $\overline{D}^*\rho$. 
The smallness of the final state interaction phase
for the two-body decay was advocated by an intuitive argument based on
QCD\cite{BJ}. It is actually required by the phase-amplitude dispersion
relation unless the amplitude is abnormally enhanced or suppressed in
magnitude\cite{Suzuki}. The possibility that the highly suppressed 
two-body decay amplitudes such as $A_{00}$ can have large decay phases 
has been predicted in the random S-matrix model of the final-state
interaction\cite{Suzuki}.  The smallness of the final-state interaction is
a phenomenon special to the two-body decay. It does not imply the same 
in the multibody or inclusive decays.  In the decays where more than 
two hadrons is produced from a heavy particle, the phase of the decay 
amplitude depends on the invariant subenergies in the final state. 
It is almost obvious theoretically that if one or more of the subenergies 
is small, the phase of the decay amplitude can be large.

To summarize, we have quantified the smallness of the final-state interaction
phases which is implied by the color suppression in the $B$ decay. According 
to the latest world-average data, the final-state interaction phases have
already been bounded fairly tightly. Our analysis shows that the current bounds 
on the color-suppressed neutral modes should not be far from their
actual values.  Lowering the upper bounds on the branching fractions of the
color-suppressed modes together with more accurate measurement of the
color-favored modes will set even severe limits on the final-state 
interaction phases or give their actual values.  They will have an important
implication in the CP violation search through the modes such as
$B\rightarrow\pi\pi$.

\acknowledgements

This work is supported in part by the Director, Office of Energy Research, 
Office of High Energy and Nuclear Physics, Division of High Energy Physics 
of the U.S.  Department of Energy under Contract DE--AC03--76SF00098 and in
part by the NSF under grant PHY--95--14797.

\begin{table}
\caption{The decay amplitudes extracted from the data.  $A_{0+}$, $A_{-+}$,
and $2^{1/2}A_{00}$ denote $A(B^+\rightarrow\overline{D}^0\pi^+)$,
$A(B^0\rightarrow D^-\pi^+)$, and 
$2^{1/2}A(B^0\rightarrow\overline{D}^0\pi^0)$, 
respectively, in the case of $B\rightarrow\overline{D}\pi$, 
and the corresponding amplitudes in other cases.
$|A_{0+}|$ is normalized to unity up to an experimental error.}

\begin{tabular}{|c|c|c|c|} 
Decay modes&     $|A_{0+}|$ & $|A_{-+}|$ & $2^{1/2}|A_{00}|$ \\ \hline
$\overline{D}\pi$ & $1\pm 0.0487$& $0.7741\pm 0.0526$ &
                      $<0.2188\pm 0.0028$\\ \hline
$\overline{D}\rho$& $1\pm 0.0682$ & $ 0.7907\pm 0.0708$ & 
                      $<0.2481\pm 0.0032$\\ \hline
$\overline{D}^*\pi$& $1\pm 0.0451$ & $0.7976\pm 0.0320$ &
                      $<0.4497\pm 0.0058$\\ \hline
$\overline{D}^*\rho$&$1\pm 0.1007$ & $0.6765\pm 0.1668$ &
                      $<0.2765\pm 0.0035$\\
\end{tabular}
\label{table:1}
\end{table}

\begin{table}
\caption{The bounds on the phases $\delta_{-+}=\arg(A_{-+}/A_{0+})$,
$\delta_{00}=\arg(A_{00}/A_{0+})$, and $\delta_I=\arg(A_{1/2}/A_{3/2})$. 
We have chosen as $0^{\circ}<\delta_{-+}<180^{\circ}$, which leads to 
$\delta_{00}<0$ and $\delta_I>0$.}

\begin{tabular}{|c|c|c|c|}
Decay modes &  $\delta_{-+} (>0)$ & $\delta_{00} (<0)$
         & $\delta_I (>0)$\\ \hline
$\overline{D}\pi$ & $<11^{\circ}$ & $>-44^{\circ}$
                  & $< 19^{\circ} $ \\ \hline
$\overline{D}\rho$ & $<16^{\circ}$ & $>-60^{\circ}$ 
                  & $< 26^{\circ} $ \\ \hline
$\overline{D}^*\pi$ & $<29^{\circ}$ & $>-59^{\circ}$
                  & $< 46^{\circ} $ \\ \hline
$\overline{D}^*\rho$ & $<21^{\circ}$ & $>-54^{\circ}$
                  & $< 40^{\circ} $ \\ \hline\hline
$D\rightarrow\overline{K}\pi$ & $80^{\circ}\pm 7^{\circ}$ & 
                $-70^{\circ}\pm 8^{\circ} $ & $90^{\circ}\pm 7^{\circ}$ \\ 
\end{tabular}
\label{table;2}
\end{table}

\noindent
\begin{figure}
\epsfig{file=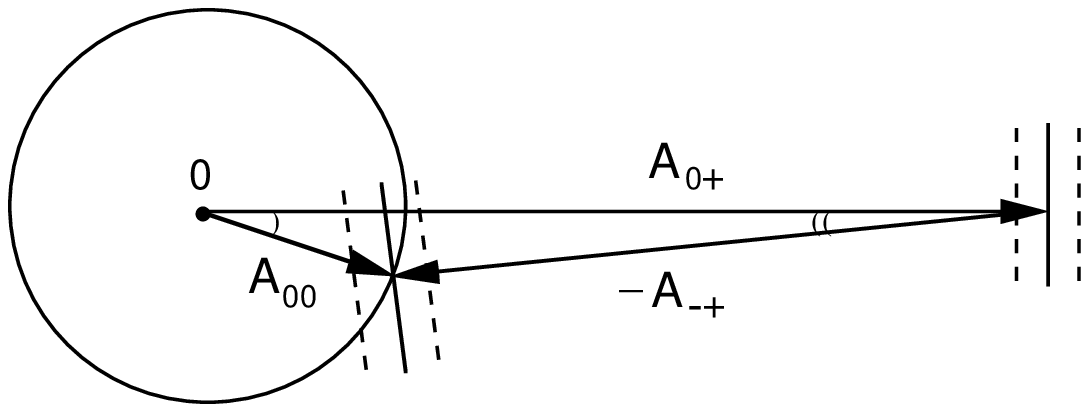,width=7cm,height=2.6cm} 
\caption{The sum rule holds in the triangular relation typically as shown here. 
The phase of $A_{0+}$ has been chosen to be zero for reference. $A_{00}$ 
is confined inside the circle. The bands indicated by broken lines at the 
ends of $A_{0+}$ and $-A_{-+}$ represent their experimental uncertainties. 
The upper bound on $|A_{00}|$ constrains the angle $\delta_{-+}$ between 
$A_{-+}$ and $A_{0+}$ to small values, while the phase of $A_{00}$
is subject to larger uncertainties than that of $A_{-+}$.}
\label{fig:1}
\end{figure}


\begin{references}

\bibitem{small} M. Bander, D. Silvermann, and A. Soni, Phys. Rev. Lett. 
             {\bf 43}, 242 (1979): 
                J.G\'{e}rard and W.-S. Hou, Phys. Rev. Lett. 
             {\bf 62} 855 (1991); Phys. Rev. D {\bf 43}, 2909 (1991): 
                G. Eilam and D. Wyler, Nucl. Phys. {\bf B352}, 367 (1991): 
                R. Fleischer, Z. Phys. C {\bf 58}, 438 (1993); C {\bf 62}
             81 (1994): 
                N. Deshpande and X.-G. He, Phys. Lett. B {\bf 336}, 471 (1994): 
                G. Kramer, W.F. Palmer, and H. Simma, Nucl. Phys. {\bf B328}, 
             77 (1994); Z. f. Phys. C {\bf 66}, 429 (1995).
\bibitem{BJ} J.D. Bjorken, Nucl. Phys. {\bf B} (Proc. Suppl.) {\bf 11} 325
                (1989).
\bibitem{Regge} H. Zhang, Phys. Lett. B {\bf 356}, 107 (1995): 
                B. Blok and I. Halperin, Phys. Lett. B {\bf 385}, 325 (1996): 
                B. Blok, M. Gronau, and J.L. Posner, Phys. Rev. Lett. 
             {\bf 78}, 3999 (1997): 
                A.N. Kamal and C. W. Luo, Phys. Lett. B {\bf 398}, 151 (1997):
                G.Narduli and T.N. Pham, Phys. Lett. B {\bf 391}, 165 (1997): 
                M. Gronau and J.L. Rosner, Enrico Fermi Institute Preprint, 
              hep-ph/9806348.
\bibitem{Suzuki} M. Suzuki, LBNL Preprint, hep-ph/9807414. 
\bibitem{large} J.F. Donoghue, E. Golowich, A. A. Petrov, and J.M. Soares,
                Phys. Rev. Lett. {\bf 77}, 2178 (1996).
\bibitem{PDG} Review of Particle Physics, Particle Data Group, C. Caso
                {\it et al.}, Eur. Phys. J. C {\bf 3}, 1 (1998).
\bibitem{Exp} CLEO Collaboration, M. Bishai {\it et al.}, Phys. Rev. Lett.
               {\bf 78}, 3261 (1997).

\bibitem{Watson} K.M. Watson, Phys. Rev. {\bf 95}, 228 (1954). 
              See also K. Aizu, Proc. Int, Conf. Theoret. Phys., 
              1953, Kyoto, Kyoto-Tokyo Science Council, 1954, p.200: 
              E. Fermi, Suppl. Nuovo. Cim. {\bf 2}, 58 (1955).

\end{references}
\end{document}